\documentstyle[12pt,supercite]{article}
\def\abstracts#1#2#3{{
        \centering{\begin{minipage}{4.62in}\baselineskip=13pt
        \small
        \centerline{\bf Abstract}
        \vspace*{0.2cm}                
        \parindent=0pt #1\par
        \parindent=18pt #2\par
        \parindent=15pt #3
        \end{minipage} }\par}}
\begin{document}
\vspace*{-2cm}
\hfill \parbox{5.5cm}{ 
                     Mainz preprint KOMA-96-21}\\
\vspace*{2cm}
\centerline{\LARGE \bf Softening of First-Order Phase Transition}\\[0.3cm]
\centerline{\LARGE \bf on Quenched Random Gravity Graphs}\\[0.4cm]
\vspace*{0.2cm}
\centerline{\large {\em Clive F. Baillie$^1$, Wolfhard Janke$^2$ and 
Desmond A. Johnston$^3$\/}}\\[0.4cm]
\centerline{\large {\small $^1$ Department of Computer Science,
         University of Colorado}}
\centerline{    {\small  Boulder, CO 80309, USA}}\\[0.5cm]
\centerline{\large {\small $^2$ Institut f\"ur Physik,
                    Johannes Gutenberg-Universit\"at Mainz}}
\centerline{    {\small Staudinger Weg 7, 55099 Mainz, Germany }}\\[0.5cm]
\centerline{\large {\small $^3$ Department of Mathematics, 
                    Heriot-Watt University}} 
\centerline{    {\small Edinburgh, EH14\,4AS, Scotland  }}\\[0.5cm]
\abstracts{}{
We perform extensive Monte Carlo simulations of the 10-state Potts model
on quenched two-dimensional $\Phi^3$ gravity graphs to study the effect of quenched
coordination number randomness on the nature of the phase transition, which
is strongly first order on regular lattices. The numerical data provides strong
evidence that, due to the quenched randomness, the discontinuous first-order
phase transition of the pure model is softened to a continuous transition, 
representing presumably a new universality class. This result is in striking 
contrast to a recent Monte Carlo study of the 8-state Potts model on 
two-dimensional Poissonian random lattices of Voronoi/Delaunay type, where the
phase transition clearly stayed of first order, but is in qualitative 
agreement with results for quenched bond randomness on regular lattices.
A precedent for such softening with connectivity disorder
is known: in the 10-state Potts model
on {\it annealed} $\Phi^3$ gravity graphs a continuous transition is 
also observed.
}{}
%
\thispagestyle{empty}
\newpage
\pagenumbering{arabic}
%
                     \section{Introduction}
%
Systems subject to quenched random disorder often show a completely different
behavior than in the pure case. If the pure system has a continuous phase
transition it is well known that quenched random disorder can drive the
critical behavior into a new universality class, or the transition can even
be eliminated altogether \cite{quench}. In the case of a first-order 
phase transition in the pure system the effect of quenched random disorder
can also be very dramatic. In fact, phenomenological renormalization-group 
arguments suggest the possibility for a softening to a continuous 
transition \cite{rg}.

The paradigm for testing the latter prediction is the two-dimensional (2D) 
$q$-state Potts model. This model is exactly known \cite{wu} to exhibit
on regular lattices for $q \ge 5$ a first-order transition whose strength 
increases with $q$. In Ref.~\cite{chen} the effect of quenched {\em bond}
disorder
was investigated for the 8-state model. By means of extensive Monte Carlo (MC)
simulations the predicted softening was confirmed, and a finite-size
scaling (FSS) analysis showed that the critical behavior of the quenched model 
could be well described by the Onsager
Ising model universality class \cite{chen}. 

In Refs.~\cite{javi95,javi95b} the effect of quenched 
{\em connectivity} 
disorder
was studied by putting the $8$-state model on 2D Poissonian random lattices
with toroidal topology, constructed according to the Voronoi/Delaunay 
prescription \cite{ranlat}. Here the bonds are all of equal strength, but the
distribution of coordination numbers ($=3,\dots,\infty$) varies randomly from
lattice to lattice, giving rise to the quenched random disorder. MC 
simulations combined with FSS analyses provided clear evidence that for this
type of quenched disorder the transition stays first order as in the pure
case.

A different, stronger\footnote{We
provide some justification for the use of ``stronger''
in what follows.}, sort of coordination number
randomness appears in 2D gravity triangulations
or their dual $\Phi^3$ graphs. In such models one is 
interested in the coupling of matter to 2D gravity, so the disorder
is annealed rather than quenched. 
Motivated by Wexler's mean field results for $q=\infty$ Potts models
coupled to 2D gravity \cite{Wex}, simulations
of the 10-state and 200-state Potts model coupled
to 2D gravity (i.e., on an annealed ensemble of $\Phi^3$ graphs of
spherical topology) gave convincing evidence \cite{2DG} for a continuous 
transition, with the measured 
critical exponents for the 10-state Potts model being consistent
with the KPZ exponents of the 4-state Potts model coupled to 2D gravity.
 
As the only quenched connectivity disorder
seriously investigated to date,
2D Poissonian random lattices,
showed no sign of softening for first-order 
transitions it is interesting to enquire
whether the salient feature for the softening in the 2D gravity
simulations is the annealed nature 
of the connectivity disorder or whether
it is some intrinsic features of the graphs themselves.
The simulations also touch on the question
of what constitutes the universality class of the
KPZ exponents for matter coupled to 2D gravity.
Recent work has suggested that the full 2D gravity
curvature distribution is not required on dynamical
triangulations in order to stay within the KPZ universality
class \cite{BB,BB1}. The current work 
can also be viewed as addressing the question of 
whether quenching the dynamical lattices has any effect 
on the exponents. 

In this note we present results of a MC study 
on quenched random lattices drawn from the equilibrium distribution
of pure 2D gravity triangulations. To be precise we simulated the 10-state
Potts 
model on the dual of these triangulations, the so-called $\Phi^3$ graphs of
spherical topology. As our main result we obtain strong evidence that for the
gravity type of random lattices the transition is indeed softened and 
seems 
to define a new ``quenched'' universality class.
%
                     \section{Model and simulation}
%
We used the standard definition of the $q$-state Potts model, 
\begin{equation}
Z_{\rm Potts} = \sum_{\{\sigma_i\}} e^{-\beta E}; 
E = -\sum_{\langle ij \rangle}
\delta_{\sigma_i \sigma_j}; \sigma_i = 1,\dots,q,
\label{eq:zpotts}
\end{equation}
where $\beta=J/k_B T$ is the inverse temperature in natural units, 
$\delta$ is the Kronecker symbol, and 
$\langle ij \rangle$ denotes the nearest-neighbour bonds of random
$\Phi^3$ graphs (without tadpoles or self-energy bubbles) with $N = 250$, 
500, 1\,000, 2\,000, 3\,000, 5\,000, and 10\,000 sites. For each
lattice size we generated 64 independent replica using the Tutte algorithm
\cite{tutte}, and performed long
simulations of the 10-state model near the transition point at 
$\hat{\beta} = 2.20$, 2.20, 2,20, 2.22, 2.23, 2.24, and 2.242,
respectively, using the single-cluster update algorithm \cite{cluster}. 
After thermalization we recorded 
$500\,000$ measurements ($750\,000$ for $N=10\,000$), taken after 6, 6, 6, 
20, 15, 15, 15 clusters had been
flipped, of the energy $E$ and the magnetization 
$M = (q \, {\rm max} \{ n_i \} - N)/(q-1)$ in 64 time-series 
files, where $n_i \le N$ denotes the number of spins of ``orientation'' 
$i=1,\dots,q$ in one lattice configuration. 
Obviously it is sufficient to
store the integers $N/q \le {\rm max} \{ n_i \} \le N$. 
The corresponding 
quantities per site are denoted in the following by $e = E/N$ and $m = M/N$.

Given this raw data we employed standard reweighting techniques \cite{rew} 
to compute, e.g., the specific heat, 
$C^{(i)}(\beta) = \beta^2 N \left( \langle e^2 \rangle - 
\langle e \rangle^2 \right)$,
for each replica labeled by the superindex $(i)$, and then performed the 
replica average, $C(\beta) \!=\! [C^{(i)}(\beta)] \!\equiv (1/64) \sum_i^{64} 
C^{(i)}(\beta)$, denoted by the square brackets. To perform the 
replica average at the level of the $C^{(i)}$ (and {\em not} at the level
of energy moments) is motivated by the general rule that quenched averages
should be performed at the level of the free energy and not the partition
function \cite{binder_young}. Finally, we
determined the maximum, $C_{\rm max} = C(\beta_{C_{\rm max}})$, for each
lattice size and studied the FSS behavior of $C_{\rm max}$
and $\beta_{C_{\rm max}}$. The error bars on the two quantities entering
the FSS analysis are estimated by jack-kniving \cite{jackknife} over the 
64 replica. This takes into account the statistical errors on the estimates
of each 
$C^{(i)}(\beta)$ as well as the
fluctuations among the different $C^{(i)}(\beta)$, $i=1,\dots,64$, caused 
by the quenched 
connectivity 
disorder of the randomly chosen $\Phi^3$ graphs.

The analysis of the magnetic susceptibility, $\chi(\beta) =
\beta N \left( [\langle m^2 \rangle - \langle m \rangle^2 ] \right)$ 
and the energetic Binder parameter\footnote{The subindex refers to the
specific choice of replica average as usually employed in spin glass
simulations. For the other options, see Ref.~\cite{javi95}.}, 
$V_3(\beta) = 1 - \left[ \langle e^4 \rangle \right]/3 
\left[ \langle e^2 \rangle \right]^2$,
proceeds exactly along the same lines, yielding $\chi_{\rm max}$ and
$\beta_{\chi_{\rm max}}$ as well as $V_{3,{\rm min}}$ and
$\beta_{V_{3,{\rm min}}}$.  
In order to be prepared for the possibility of a second-order phase
transition, the magnetic Binder parameter,
$U_3(\beta) = 1 -
\left[ \langle m^4 \rangle \right] /3 \left[ \langle m^2 \rangle \right]^2$,
was also measured as, in this case, its crossing 
for different lattice sizes provides an alternative
determination of the critical coupling $\beta_c$, and the 
FSS
of either the maximum slopes or the slopes at 
$\beta_c$ can be used to extract the correlation length exponent $\nu$.
%
                     \section{Results}
%
The maxima of the specific heat and the susceptibility are shown in
Fig.~1 on a linear scale. If the transition was of first-order one would
expect for large system sizes an asymptotic FSS behavior of the 
form $C_{\rm max} = a_C + b_C N + \dots$, and $\chi_{\rm max} = a_{\chi} + 
b_{\chi} N + \dots$ \cite{scaling}. As is evident from Fig.~1
a linear scaling with $N$ is clearly not consistent with our data. Even though
it is intrinsically never clear what could happen for much larger lattice
sizes, we take this as evidence for the softening to a continuous phase
transition. Furthermore, 
at a first-order phase transition one would expect that
the energetic Binder-parameter minima approach 
in the infinite-volume limit a non-trivial value 
related to the latent heat.  
Our data shown in Fig.~2, however, approach
the trivial limit of $2/3$ which is another firm indication that the 
transition of the 10-state Potts model on quenched 2D gravity graphs is 
{\em not\/} of first-order.

Being thus convinced that the transition is continuous, the next goal is
to determine the critical exponents and the corresponding universality class.
To this end we have tried to describe the scaling of the specific-heat and 
susceptibility maxima with the standard FSS ansatz
\begin{equation}
  C_{\rm max} = a_C + b_{C} N^{\alpha/D \nu},
  \label{eq:fss_C}
\end{equation}
and
\begin{equation}
  \chi_{\rm max} = a_{\chi} + b_{\chi} N^{\gamma/D \nu},
  \label{eq:fss_chi}
\end{equation}
where $\alpha$, $\gamma$, and $\nu$ are the usual universal critical 
exponents at a continuous phase transition, $a_{C,\chi}$ and $b_{C,\chi}$ are
non-universal amplitudes, and $D$ is the intrinsic Hausdorff dimension of
the graphs.

By performing a non-linear three-parameter fit to the specific-heat maxima
we obtained $\alpha/D \nu = 0.22(7)$, with a reasonable goodness-of-fit 
parameter $Q=0.10$. Since the background term $a_C$ turned out to be
consistent with zero, we also tried linear two-parameter fits with 
$a_C=0$ kept fixed. By omitting successively the smaller lattice sizes
the resulting exponent estimates varied in the range of
$\alpha/D \nu = 0.21 \dots 0.26$.
In particular the fit over all data points gave a fully
consistent value of $\alpha/D \nu = 0.222(7)$ (with $Q=0.15$) but, 
as expected, a much smaller error bar.
%
%
The susceptibility maxima grow very fast with $N$, such that also here the
constant term $a_{\chi}$ can safely be neglected. The linear fit over
all data points yielded $\gamma/D\nu = 0.732(10)$ (with $Q=0.27$), and
omitting the $N=250$ point we obtained $\gamma/D \nu = 0.719(14)$ (with
$Q=0.31$). The fits are shown together with the data on the log-log scale
of Fig.~3. 

The pseudo-transition points $\beta_{C_{\rm max}}$, $\beta_{\chi_{\rm max}}$,
and $\beta_{V_{3,{\rm min}}}$ are shown in Fig.~4 together with fits to
the standard FSS ansatz   
\begin{equation}
\beta_{C_{\rm max}} = \beta_c + c_C N^{-1/D\nu},
\label{eq:beta_scal}
\end{equation}
etc. Here we
fitted again all available
data down to $N=250$ and in this way obtained at least for $\beta_c$
stable estimates. Our results are collected in Table~\ref{tab:bc_fits}, 
where we also give the exponent $1/D \nu$ which, however, has already quite 
large errors. By taking the average of the three estimates of $\beta_c$ in 
Table~\ref{tab:bc_fits}, we arrive at our final estimate of the transition 
point,
\begin{equation}
  \beta_c = 2.2445 \pm 0.0020.
  \label{eq:bc}
\end{equation}
This value is consistent with the estimate obtained from the crossings
of the magnetic Binder parameter $U_3$ for different lattice sizes.
\begin{table}
\caption{Fit results of $\beta_c(N) =\beta_c + a N^{-1/D\nu}$
to the pseudo-transition points. }
\begin{center}
\begin{tabular}{|l|l|l|l|}
\hline
\multicolumn{1}{|c|}{obs.} & \multicolumn{1}{c|}{$\beta_c$} &
\multicolumn{1}{c|}{$1/D\nu$} & \multicolumn{1}{c|}{$Q$} \\
\hline
$\beta_{C_{\rm max}}$     & 2.2448(29) & 0.56(13)  & 0.14 \\
$\beta_{\chi_{\rm max}}$  & 2.2450(17) & 0.88(54)  & 0.14 \\
$\beta_{V_{3,{\rm min}}}$ & 2.2435(20) & 0.74(14) & 0.67 \\
\hline
\end{tabular}
\end{center}
\label{tab:bc_fits}
\end{table}
It is interesting to note that this value (allowing
for a factor of two in the normalization of $\beta$)
is very close to the $\beta_c$ measured for the 10-state
Potts model on annealed 2D gravity graphs in \cite{clivedes}
\footnote{The transition point measured in
\cite{2DG} is different, but in this case
tadpoles were allowed in the ensemble of graphs.},
namely $\beta_c = 1.141(1)$. A similar effect is apparent
in the simulations of the Ising model on quenched graphs
in \cite{bhj94} where the measured $\beta_c$ is identical
to the annealed graph value.

A more precise estimate of the exponent $1/D\nu$ can be obtained by
analyzing the FSS of the magnetic Binder-parameter slopes in the
vicinity of $\beta_c$. Both, the maximum slopes and the slopes at
$\beta_c$, are expected to scale as $d U_3/d\beta \propto N^{1/D\nu}$. 
The fit to the maximum slopes for $N \ge 500$
gave $1/D\nu = 0.616(29)$ (with $Q=0.29$), and for the slopes at 
$\beta_c=2.2445$ we obtained $1/D\nu = 0.614(30)$ (with $Q=0.78$).

We now take stock of our numerical results. For convenience
we show in Tables~2 and 3 the analytically calculated
critical exponents for the 2-state Potts (Ising) and
4-state Potts model on flat 2D lattices, 
coupled to annealed 2D gravity (the standard KPZ result),
and on quenched 2D gravity graphs (calculated
in \cite{quench_theory} by taking $n$ matter copies with $n
\rightarrow 0$ in the KPZ formulae)

\vskip 0.5cm

\centerline{Table 2:  Critical exponents for Ising models}
\begin{center}
\begin{tabular}{|c|c|c|c|c|c|c|c|c|c|} \hline
Type& $\alpha /D \nu$  & $\gamma /D \nu$  & $1 /D \nu$ \\[.05in]
\hline
Fixed& $0$  & $7/8$ & $1/2$ \\[.05in]
\hline
Annealed& $-1/3$  & $2/3$ & $1/3$ \\[.05in]
\hline
Quenched& $-0.303...$  & $0.709...$ & $0.349...$  \\[.05in]
\hline
\end{tabular}
\end{center}
\vspace{.1in}

\vskip0.8cm

\centerline{Table 3:  Critical exponents for $q=4$ Potts models}
\begin{center}
\begin{tabular}{|c|c|c|c|c|c|c|c|c|c|} \hline
Type& $\alpha /D \nu$  & $\gamma /D \nu$  & $1 /D \nu$ \\[.05in]
\hline
Fixed& $1/2$  & $7/8$ & $3/4$ \\[.05in]
\hline
Annealed& $0$  & $1/2$ & $1/2$ \\[.05in]
\hline
Quenched& $0.177...$  & $0.709...$ & $0.589...$  \\[.05in]
\hline
\end{tabular}
\end{center}
\vspace{.1in}

\vskip0.5cm

Given the measured value 
of $\gamma /D \nu \simeq
0.72(2)$ it might seem tempting to conclude, similar to the random bond
case, that the critical behavior can be related to an Ising universality class.
While our estimate for $\gamma/D \nu$ is very far off the Onsager value of
$\gamma/D \nu = 0.875$ for 2D regular lattices, it is remarkably close to 
the value for an Ising model on {\em annealed\/} gravity graphs, 
$\gamma/D \nu = 0.666\dots$, and even fully compatible with the theoretical 
prediction for the Ising model on quenched $\Phi^3$ graphs
\footnote{As $\gamma/D \nu$ depends
only on the conformal weight of the spin operator
on both fixed and quenched lattices
the values for the Ising and 4-state Potts
model are ``accidentally'' equal.}, 
$\gamma/D \nu = 0.709\dots$ \cite{quench_theory}. 
However, 
the same value is also predicted for the quenched 4-state Potts model and
the estimates for $\alpha /D \nu \simeq 0.22(1)$
and $1/D \nu \simeq 0.61(3)$ itself are closer to the 
critical exponents of the
4-state Potts model on quenched 2D gravity graphs listed in the bottom line
of Table 3. 
The results are {\it not} consistent with the KPZ critical
exponents of the 4-state Potts model on annealed graphs  because
of the relatively large difference 
(compared with the Ising case) between the quenched
and annealed exponents for $q=4$.

On the evidence of our estimates, it would thus seem that the 
critical exponents of the 10-state Potts model
on quenched 2D gravity graphs are best fitted by those
of the 4-state Potts model on quenched 2D gravity graphs
as listed in Table 3.
The situation appears to be entirely analogous to that
for the 10-state Potts model on annealed 2D gravity graphs, where
the measured critical exponents are consistent with 
the annealed (KPZ) exponents for the 4-state Potts model
coupled to 2D gravity.
%
                     \section{Conclusions}
%
To summarize, we have obtained strong numerical evidence that 
due to connectivity disorder 
the phase 
transition in the 10-state Potts model on quenched random gravity graphs
is softened to a continuous phase transition. This result is in contrast to
a recent simulation of the 8-state model on Poissonian Delaunay/Voronoi
random 
lattices where the transition stays first order as on regular 
lattices \cite{javi95,javi95b}. 
It is, however, in qualitative agreement with the quenched 
random bond case \cite{chen},
and in particular it does follow the pattern of \cite{bhj94} where the
Ising transition on a quenched ensemble of 2D gravity graphs 
was of similar order to its counterpart on annealed graphs.
Like the simulations in \cite{2DG} of the 10-state Potts model on
annealed 2D gravity graphs we 
see exponents associated with the 4-state Potts model,
in this case the ``quenched'' exponents. 
An obvious extension of the current work is to look
at the 200-state Potts model on quenched graphs to see
if, analogously to \cite{2DG},
Ising exponents are obtained
(presumably the quenched exponents rather than the KPZ
exponents in this case), and to study the associated cross-over as a
function of $q$.
As the current work provides rather stronger evidence for
the quenched exponents than the Ising simulation
of \cite{bhj94} it would also be a worthwhile exercise
to improve the statistics of this work to pin
down the (small) difference between the annealed and quenched
Ising exponents in the measurements.

As a parenthetical remark, 
it is worth noting that we have carried out our simulations
on an ensemble of {\it pure} 2D gravity graphs, as this seemed
the most natural choice from the theoretical considerations,
where the back reaction of the matter
on the geometry is essentially switched off when taking the quenched
limit. It is not clear what to expect if one simulates
the models on a quenched ensemble with the ``wrong''
background charge, such as $c=-2$. Simply switching off
the back reaction in an annealed simulation leads to
instability as there is insufficient separation between the
graph and spin model time scales in such a case.  

Without further theoretical work it is presumably very difficult to
isolate the distinguishing parameters that classify the disorder in
the coordination numbers as ``strong'' (gravity graphs) and ``weak''
(Poissonian lattices). 
One might think
that the very different coordination number distributions
for Poissonian Voronoy/Delaunay lattices 
and random gravity triangulations (the dual of $\Phi^3$ graphs) 
as shown in Fig.~5 were an important factor,
but \cite{BB1} carried out a fairly brutal
truncation of the 2D gravity distribution and still
found the KPZ exponents for an Ising transition.
It is possible that some other geometrical features
are the deciding factor:
by Euler's relation, for both types of random lattices
the average coordination number is 6 (up to a $1/N$ correction for spherical
topology), but the higher moments are obviously very different. Long-range 
correlations in the lattice geometry, however, play presumably an even more
important role than the local moments. As a first step in this direction it
would be interesting to compare nearest-neighbour correlations as described by
Aboav's or Weaire's law for the two different random lattice geometries
\cite{foam}.

The current work also throws an interesting light on \cite{BB},
where it was observed that fluctuating connectivity in a 
{\it flat} 2D geometry was sufficient to obtain the KPZ exponents
for the Ising model which are normally presumed to be 
produced by the curvature fluctuations of 2D gravity.
In our simulations we have found the softening effects of 2D
gravity persisting even when there is no fluctuating 
connectivity during the course of a simulation, even
though we appear to have obtained a different, quenched set of exponents. 
The question of
exactly what characterizes the universality class of the 
modified 2D gravity exponents 
(fluctuating connectivities, graph geometry,\dots ?), 
whether annealed or quenched, 
is still open.

%
%
\section*{Acknowledgements}
%
%
This work was supported in part by EC HCM network grant 
ERB-CHRX-CT930343.
C.F.B. and D.A.J. were supported in part by NATO collaborative
research grant CRG951253.
W.J. thanks the Deutsche Forschungsgemeinschaft for a Heisenberg fellowship.
The Monte Carlo simulations were performed on a T3D parallel computer
of Konrad-Zuse-Institut Berlin. 
%
%

\newpage
%
%
%
\begin{figure}[bhp]
\vskip 6.5truecm
\includegraphics{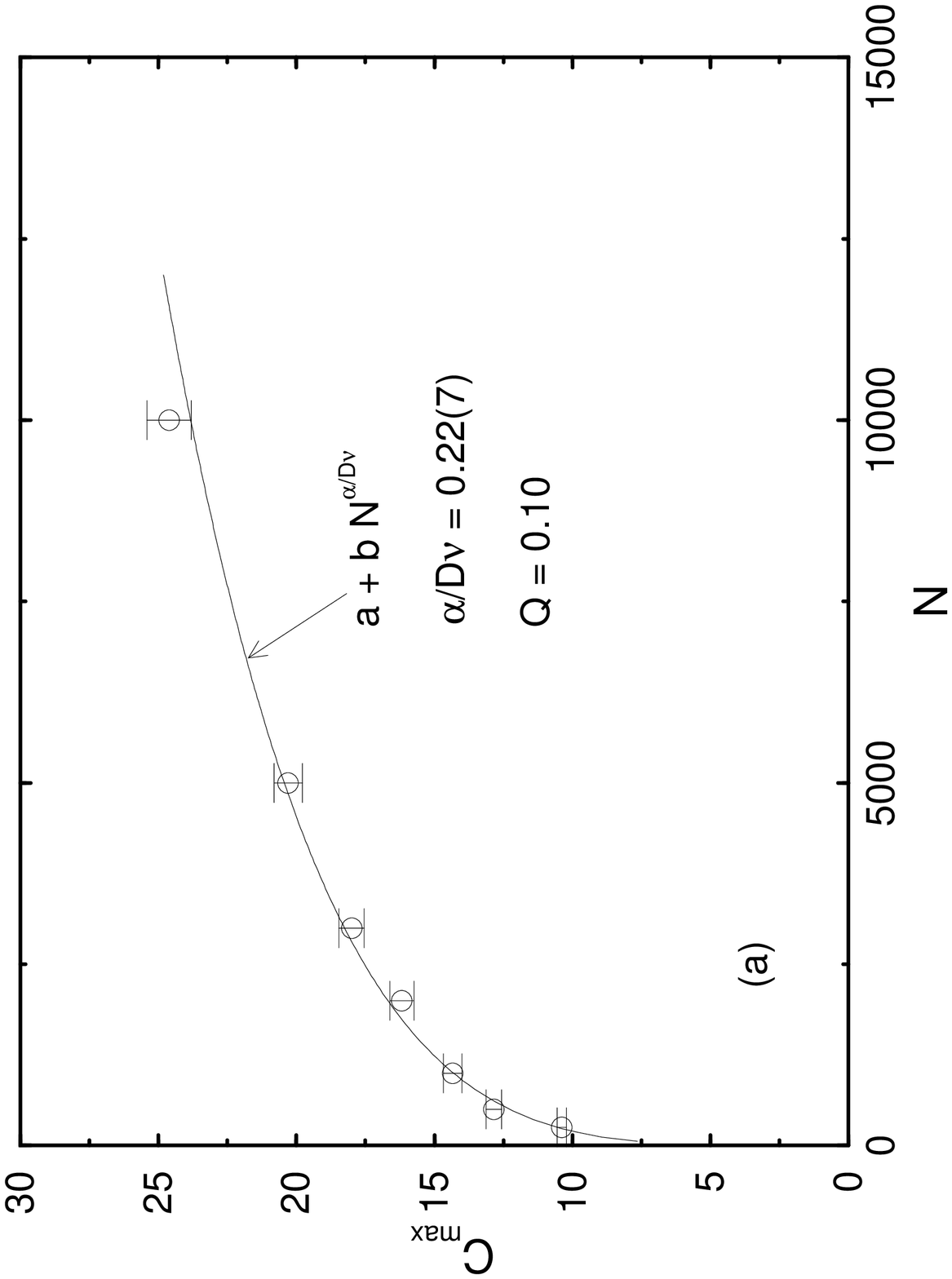}
\label{fig:c_all}
\end{figure}
\begin{figure}[bhp]
\vskip 7.5truecm
\includegraphics{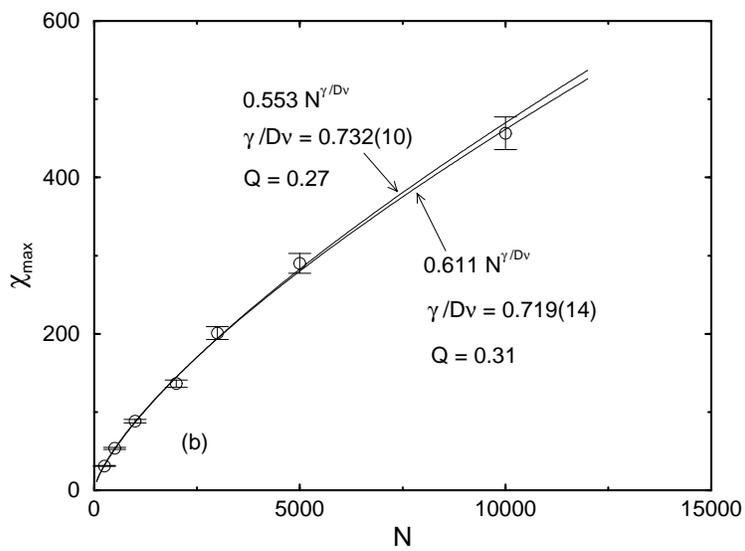}
\caption[a]{The specific-heat (a) and susceptibility (b) maxima vs
the number of sites, $N$, showing that the data are {\em incompatible\/}
with a scaling $\propto N$, which would be characteristic for a 
first-order phase transition.}
\label{fig:c_chi_max}
\end{figure}
\clearpage\newpage
%
%
\begin{figure}[bhp]
\vskip 8.0truecm
\includegraphics{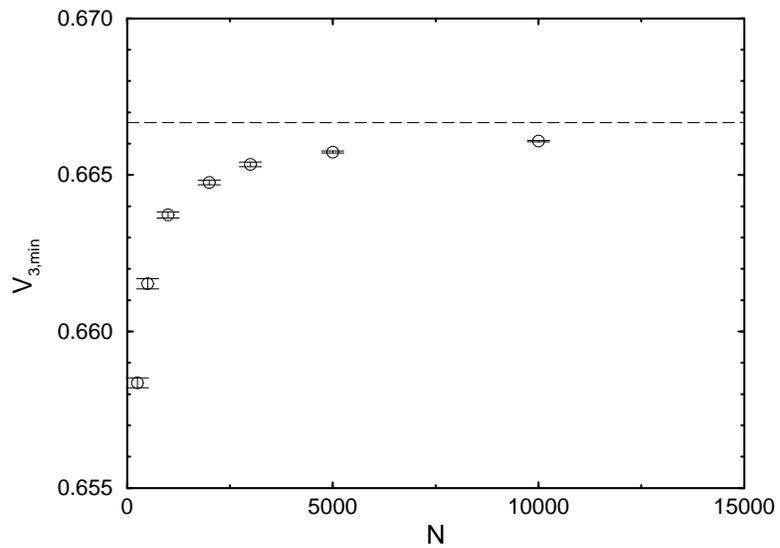}
\caption[a]{Finite-size scaling of the energetic Binder-parameter minima. 
The horizontal dashed line shows the trivial limit $2/3$.} 
\label{fig:bind_min}
\end{figure}
\clearpage\newpage
%
%
\begin{figure}[bhp]
\vskip 6.5truecm
\includegraphics{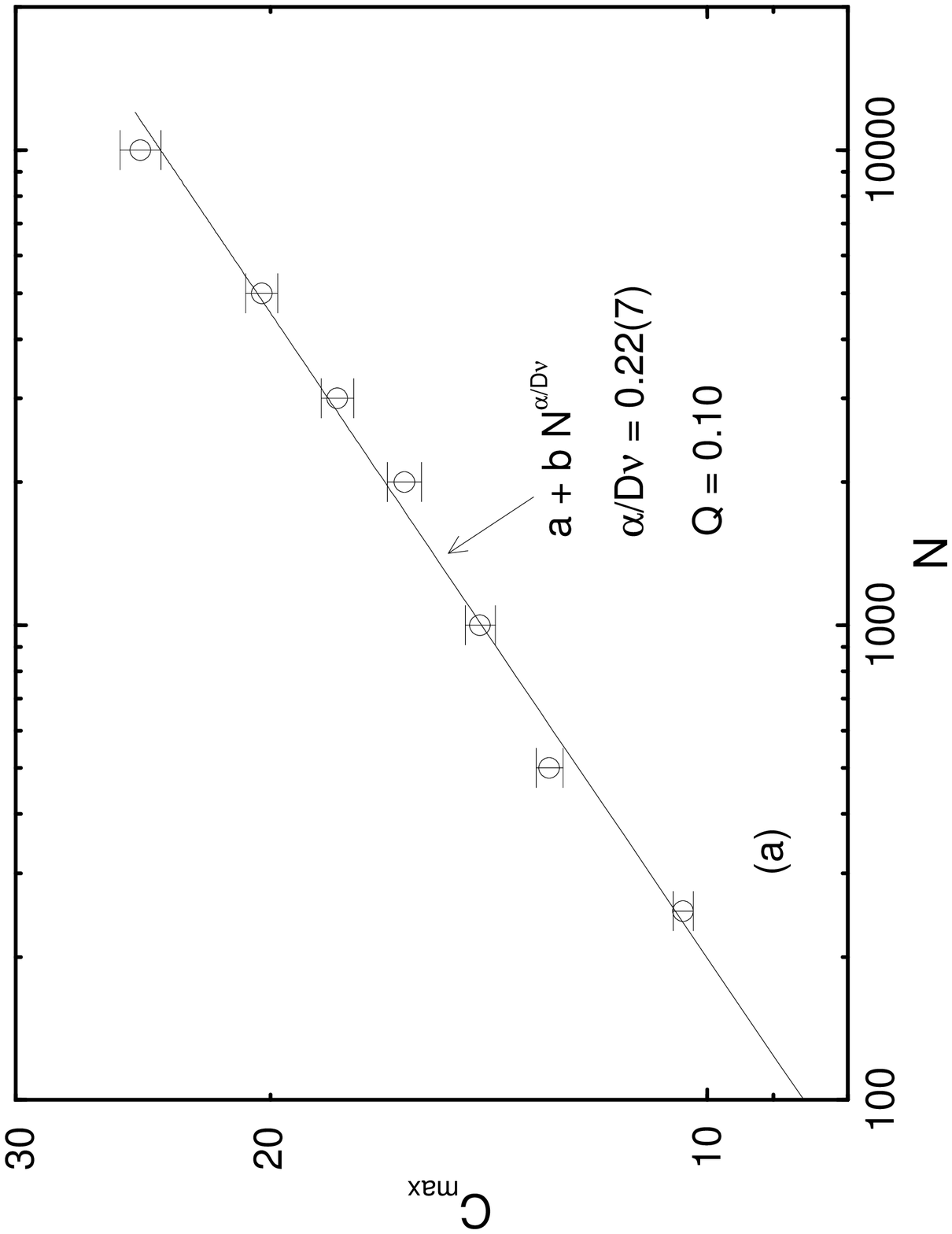}
\label{fig:c_all2}
\end{figure}
\begin{figure}[bhp]
\vskip 7.5truecm
\includegraphics{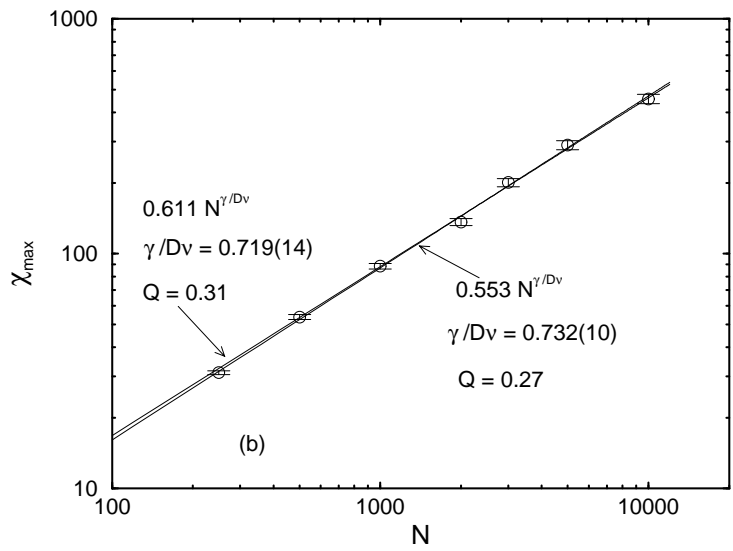}
\caption[a]{Finite-size scaling of (a) specific-heat and (b) susceptibility
maxima on a log-log scale, together with the fits discussed in the text.}
\label{fig:c_chi_max_log}
\end{figure}
\clearpage\newpage
%
%
\begin{figure}[bhp]
\vskip 7.5truecm
\includegraphics{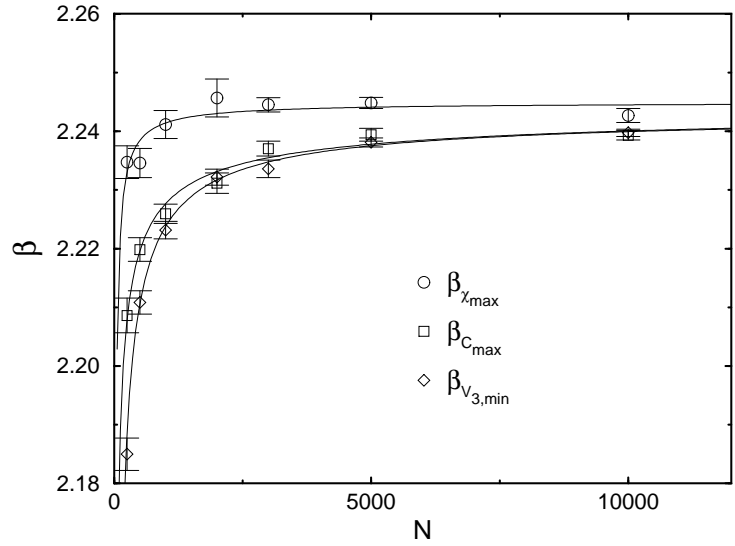}
\caption[a]{Finite-size scaling of the pseudo-transition points,
together with the fits discussed in the text.}
\label{fig:beta}
\end{figure}
%
%
\begin{figure}[bhp]
\vskip 7.5truecm
\includegraphics{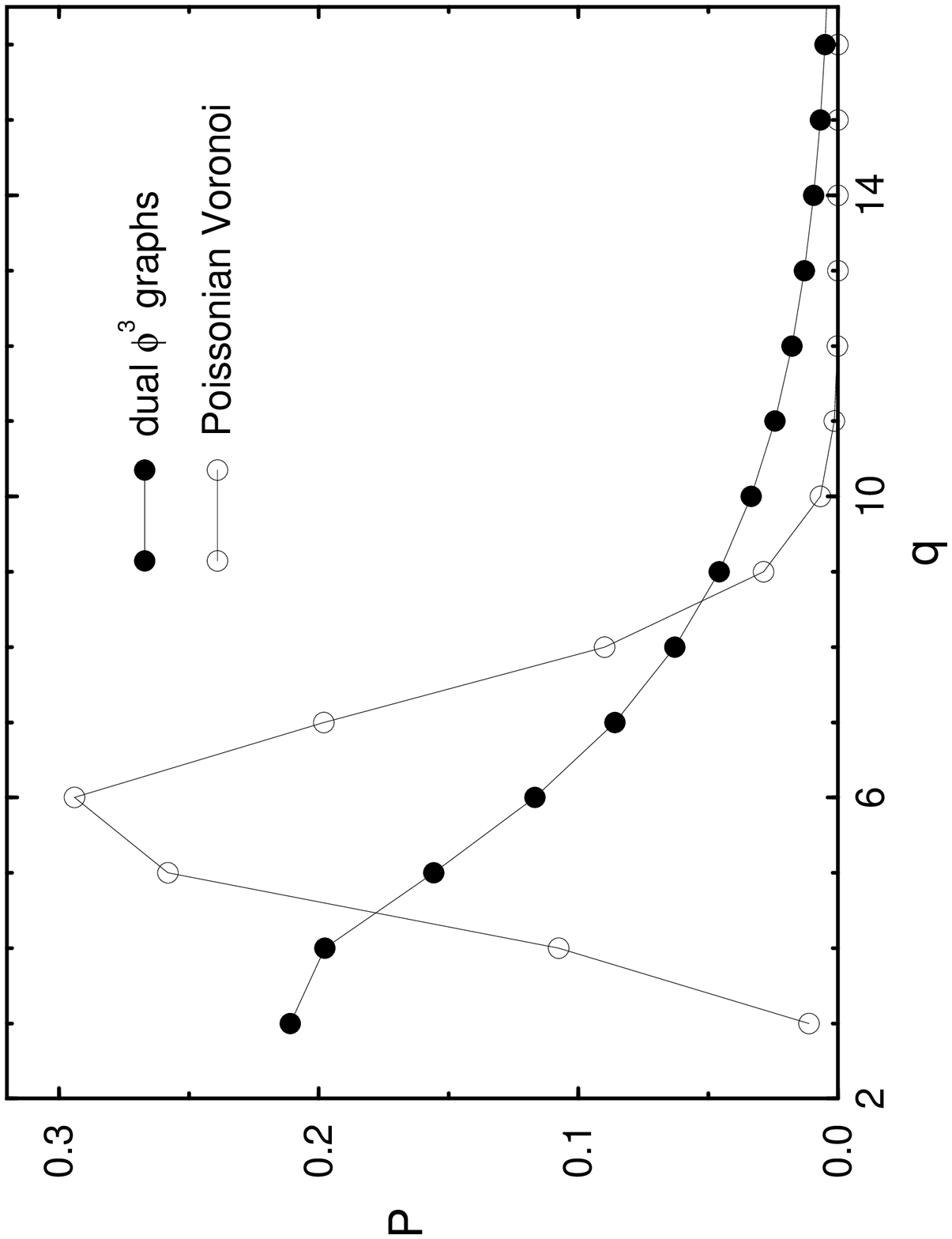}
\caption[a]{Distribution of coordination numbers for gravity triangulations
dual to $\Phi^3$ graphs and Poissonian
Voronoi/Delaunay random lattices.}
\label{fig:distrib}
\end{figure}

\begin{thebibliography}{99}
%
%
\bibitem{quench}
A.B. Harris, J. Phys. {\bf C7} (1974) 1671;
Y. Imry and S.-k. Ma, Phys. Rev. Lett. {\bf 35} (1975) 1388;
A. Aharony, Phys. Rev. {\bf B18} (1978) 3318;
G. Grinstein and S.-k. Ma, Phys. Rev. Lett. {\bf 49} (1982) 684;
A.N. Berker, Phys. Rev. {\bf B29} (1984) 5293.
%
\bibitem{rg}
A.N. Berker, Phys. Rev. {\bf B29} (1984) 5293;
K. Hui and A.N. Berker, Phys. Rev. Lett. {\bf 62} (1989) 2507; {\em ibid.\/}
{\bf 63} (1989) 2433(E); A. Aizenman and J. Wehr, Phys. Rev. Lett. {\bf 62} 
(1989) 2503; A.N. Berker and K. Hui, in {\em Science and Technology of
Nanostructured Magnetic Materials\/}, eds. G.C. Hadjipanayis, G. Prinz,
and L. Paretti (Plenum, New York, 1991).
%
\bibitem{wu}
F.Y. Wu, Rev. Mod. Phys. {\bf 54} (1982) 235; {\em ibid.\/} 
{\bf 55} (1983) 315(E).
%
\bibitem{chen}
S. Chen, A.M. Ferrenberg, and D.P. Landau,
Phys. Rev. Lett. {\bf 69} (1992) 1213; 
Phys. Rev. {\bf E52} (1995) 1377.
%
\bibitem{javi95}
W. Janke and R. Villanova, Phys. Lett. {\bf A209} (1995) 179.
%
\bibitem{javi95b}
W. Janke and R. Villanova, 
Nucl. Phys. {\bf B} (Proc. Suppl.) {\bf 47} (1996) 641.
%
\bibitem{ranlat}
C. Itzykson and J.-M. Drouffe, {\em Statistical Field
Theory\/} (Cambridge University Press, Cambridge, 1989), Vol.~2;
N.H. Christ, R. Friedberg, and T.D. Lee, Nucl. Phys. {\bf B202} (1982) 89;
Nucl. Phys. {\bf B210} [FS6] (1982) 310, 337;
R. Friedberg and H.-C. Ren, Nucl. Phys. {\bf B235} [FS11] (1984) 310.
For a recent study of the Ising model on 2D random lattices of this type
see, W. Janke, M. Katoot, and R. Villanova, Phys. Rev. {\bf B49} (1994) 9644.
%
\bibitem{Wex}M. Wexler, Phys. Lett. {\bf B315} (1993) 67;
Mod. Phys. Lett. {\bf A8} (1993) 2703; 
Nucl. Phys. {\bf B410} (1993) 477;
{\em ibid.\/} {\bf B438} (1995) 629.
\bibitem{2DG} J. Ambj\o rn, G. Thorleifsson, and M. Wexler, 
Nucl. Phys. {\bf B439} (1995) 187.
\bibitem{BB} W. Beirl and B. Berg, Nucl. Phys. {\bf B452} (1995) 415.
\bibitem{BB1} M. Bowick, S. Catterall, and G. Thorleifsson, 
``Minimal Dynamical Triangulations
of Random Surfaces'', preprint hep-th/9605167.
\bibitem{tutte}
M.E. Agishtein and A.A. Migdal, Nucl. Phys. {\bf B350} (1991) 690.
%
\bibitem{cluster}
R.H. Swendsen and J.S. Wang, Phys. Rev. Lett. {\bf 58} (1987) 86;
U. Wolff, Phys. Rev. Lett. {\bf 62} (1989) 361.
%
\bibitem{rew}
A.M. Ferrenberg and R.H. Swendsen, Phys. Rev. Lett. {\bf 61} (1988)
2635; {\em ibid.\/} {\bf 63} (1989) 1658(E).
%
\bibitem{binder_young}
K. Binder and A.P. Young, Rev. Mod. Phys. {\bf 58} (1986) 801.
%
\bibitem{jackknife}
R.G. Miller, Biometrika {\bf 61} (1974) 1; B. Efron,
{\em The Jackknife, the Bootstrap and other Resampling Plans\/}
(SIAM, Philadelphia, PA, 1982).
%
\bibitem{scaling}
See, e.g., W. Janke, in {\em Computer Simulations in Condensed Matter 
Physics VII\/}, eds. D.P. Landau, K.K. Mon, and H.B. Sch\"uttler (Springer
Verlag, Heidelberg, Berlin, 1994), p.~29; and references therein.
%
\bibitem{clivedes}
C.F. Baillie and D.A. Johnston, Mod. Phys. Lett. {\bf A7} (1992) 1519.
%
\bibitem{bhj94}
C.F. Baillie, K.A. Hawick, and D.A. Johnston, Phys. Lett. {\bf B328}
(1994) 284.
%
\bibitem{quench_theory}
D.A. Johnston, Phys. Lett. {\bf B277} (1992) 405.
%
%
%
%
%
%
%
\bibitem{foam} 
C. Godr\`eche, I. Kostov, and I. Yekuteli, 
Phys. Rev. Lett. {\bf 69} (1992) 2674.
%
\end{thebibliography}
\end{document}